\newcounter{myctr}

\documentclass{ws-acs}

\usepackage{amsmath,amsfonts,amssymb}
\usepackage[all]{xy}
\usepackage{url}

\begin{document}

\makeatletter
\def\@biblabel#1{[#1]}
\makeatother

\markboth{Ihor Lubashevsky and Natalia Plawinska}
{Mathematical formalism of physics of systems with motivation}

%
\catchline{}{}{}{}{}
%

\title{MATHEMATICAL FORMALISM OF \\ PHYSICS OF SYSTEMS WITH MOTIVATION}

\author{\footnotesize IHOR LUBASHEVSKY}

\address{A.M. Prokhorov General Physics Institute, Russian Academy of Sciences, Vavilov Str. 38,
        \\Moscow, 119991, Russia\\ ialub@fpl.gpi.ru}

\author{\footnotesize  NATALIA PLAWINSKA}
\address{Philology Faculty, Moscow State Pedagogical University, Mal. Pirogovskaya str. 1,
        \\Moscow, 119882, Russia\\ n.plavi@mail.ru}

\maketitle

\begin{history}
\received{(received date)}
\revised{(revised date)}
\end{history}

\begin{abstract}

The paper discusses fundamental problems in mathematical description of social systems based on physical concepts, with so-called statistical social systems being the main subject of consideration. Basic properties of human beings and human societies that distinguish social and natural systems from each other are listed to make it clear that individual mathematical formalism and physical notions should be developed to describe such objects rather then can be directly inherited from classical mechanics and statistical physics. As a particular example systems with motivation are considered. Their characteristic features are analyzed individually and the appropriate mathematical description is proposed. Finally the paper concludes that the basic elements necessary for describing statistical social systems or, more rigorously, systems with motivation are available or partly developed in modern physics and applied mathematics.

\end{abstract}


\section{Introduction}

During the last years it has become evident that a novel interdisciplinary branch of science, physics of social systems or sociophysics, is currently under development (for a review of the state of art in this field and its history see, e.g., \cite{1,2,3,4,5,6}). Various social processes and phenomena observed in large groups of people integrated together by some activity have become the subject matter of this science. Voting behavior in elections, opinion formation, culture and language evolution, cooperative interaction among trade agents, dynamics of traffic and pedestrian flows, etc. are typical objects of such investigations.\footnote{It should be noted that collective behavior of animal groups, such as schools of fish, flocks of birds or swarms of insects (for a review see, e.g. \cite{animal1,animal2}) is actually a relative problem to the subject under consideration.}
In some sense, sociophysics can be regarded as a novel branch of the ``quantitative sociodynamics''. The latter deals with the mathematical modeling of various phenomena out of different sectors of the human society conventionally investigated in separate social sciences such as demography, sociology, political science, and economics (for review and discussion of basic mathematical ideas of the quantitative sociodynamics see, e.g., \cite{6.1,6.2,6.3,6.4}).

At the current state of art, the mathematical formalism of the quantitative sociodynamics and, correspondingly, sociophysics is based on the notions and methods developed in statistical physics of many-particle systems and synergetics (e.g., \cite{Haken}). The systems under consideration in this scope consist of different organizational strata and involve, in particular, a microlevel and a macrolevel as well as a ``bottom up'' and ``top down'' interaction between them. The microlevel comprises individuals with their inclinations, decisions, and actions; the macrolevel contains collective political and economic structures, social trends, etc. \cite{6.3,6.4}.

In the modern approach the mathematical description of social systems at the macrolevel is reduced, first, to finding a set of appropriate order parameters, i.e., collective material and personal variables, whose time variations specify the system dynamics completely \cite{6.3,6.4}. This point is rather similar to one in describing complex systems of inanimate nature. Then, the construction of the equations governing these order parameters can be implemented, roughly speaking, in two ways.

Within a phenomenological approach the corresponding mathematical models are developed based purely on available macroscopic empirical data. In this case the question of family of models will best describe given social systems is solely a matter of empirical work.

The other possibility is to derive the required governing equations based on the elementary dynamics of the system elements determined at the microlevel. In this feature social and inanimate systems are quite distinct from each other. Strictly speaking, there are no mathematical constructions governing individual social behavior of human beings \cite{6.3}. Besides, in contrast to physical particles the behavior of human beings is the complex outcome of many physiological and psychological processes in addition to social aspects under consideration. So any modeling of social behavior of human beings inevitably involves a huge and unwarranted simplification of the real situation \cite{1}. Nevertheless these problems can be overcome for some systems comprising many individuals (elements) similar in behavior within a certain social sector under consideration. In systems listed above in relation to sociophysics the principle of local self-averaging holds. It implies that analyzed phenomena are mainly controlled by cooperative interaction of individuals, i.e. based on cumulative contribution of many elements. As a result, the individual properties of similar elements are averaged in some manner during their interaction. The latter feature enables us, first, to introduce the notion of characteristic element (social agent) with properties being the same for all such individuals. Second, in this way the individuality of human beings can be taken into account in terms of random factors characterized by statistical properties being again the same for all the characteristic elements of one type. In addition, the effect of ``social influence'' \cite{Festinger} stimulates individuals to behave alike. Exactly the principle of local self-averaging is grounds for applying the techniques of statistical physics to describing these social systems which will be referred below as to statistical social systems. In this way dealing with a specific class of phenomena the microlevel of statistical social systems is mimicked by a statistical ensemble of the relevant characteristic elements. Moreover, we note that the feasibility of introducing the macroscopic order parameters seems to be a direct consequence of the local self-averaging.

To avoid confusion it should be pointed out that here the principle of local self-averaging is used only to introduce the notion of characteristic element and does not necessary cause the mean field description to hold. For instance, the introduction of aggregate variables in macroeconomics can fail when fluctuations in the behavior of economic agents become crucial (for review of this problem see, e.g. \cite{NSA1,NSA2,NSA3}). Nevertheless even in this case the principle of local self-averaging holds if such fluctuations comprise many individuals behaving alike.

In the present paper we focus out attention on the second way of constructing the governing equations for macroscopic order parameters. Namely, we discuss the fundamental features of constructing characteristic elements mimicking the microlevel of statistical social systems, which actually is one of the main subjects of sociophysics. The relevant mathematical models have to take into account basic peculiarities of human beings, in particular, the effects of purposefull or even irrational decision making, motivations and trends in human behavior as well as a contingent ``free will'' rooted in human nature \cite{6.4}. By way of example so-called systems with motivation will be considered in detail. It should be noted that the majority of mathematical models proposed for statistical social systems use the notions and concepts inherited directly from physics  (see, e.g., reviews \cite{1,6}). These models, however, are applicable to social systems as a rough approximation only, at least, on the ``microscopic'' (most detailed) level because of takeing into account just a few of the basic peculiarities of human beings.

In addition we would like to point out that in spite of the progress achieved in sociophysics the long-standing question of whether mathematical formalism and physical notions are applicable, at least in principle, to describing social systems is actual up to now. There are widely different opinions on this question, including the well known points of the classics of sociology. Emile Durkheim claimed that from the general point of vies social and natural disciplines are rather similar in methods and approaches \cite{Durkheim}, whereas Max Weber considered them to be fundamentally distinctive \cite{Weber}. The latter point is highly relevant to constructing the ensembles of characteristic elements for the microlevel of social systems. So discussing the main subject of the present paper we make an attempt to restate the given question and in this way to overcome this contradiction. Assuming the statistical social systems to admit of a description in terms of physical and mathematical notions we pose a question as to what mathematical formalism is appropriate and able to take into account the basic peculiarities of social systems distinguishing them from the inanimate objects.

Leaping ahead, it should be noted once more that the goal of this partly debatable paper is not to construct a specific model for a certain particular phenomenon or system. By this paper we would like to attract the attention of the physical society to the question what novel physical notions and mathematical formalism should be developed or used in constructing relevant models for statistical social systems at the microlevel. Human beings are so different from objects of inanimate nature in their properties that the necessity of such investigation seems to be actually ``self-evident''. To elucidate this point of view let us first list some of the basic peculiarities of social objects distinguishing them clearly from systems of inanimate nature.

\section{Peculiarities of social systems}\label{sec:pecul}

Below we will list basic characteristics distinguishing social and physical systems from each other that are essential for the further mathematical constructions.

\begin{itemize}
\item (\textit{Individuality and complexity}) Social systems are made up of elements (individuals, agents, decision makers, etc.) with pronounced individuality in behavior and cognition. Besides, as noted in the Introduction, from the standpoint of the modeling, human beings are maltifactorial objects; the detailed behavior of each of them is a complex outcome of not only social processes but a large number of physiological and psychological ones \cite{1}. So many factors really influencing social phenomena remain uncontrollable and lying beyond the analysis of social systems. By contrast, elements of physical systems are assumed to admit of a complete description at least on the ``microscopic'' level and elements of one type are identical in properties.

\item(\textit{Uncertainty}) The individuality and complexity of human beings endow social systems with original variability and partial uncertainty. As a result both the regular and random factors are present in their dynamics on any level of detail \cite{6.3,6.4}. The classical dynamics of natural objects is deterministic on the ``microscopic'' level and the probabilistic formalism used in the statistical physics is caused by their reduced description.\footnote{The quantum uncertainty reflecting the wave-particle dualism is unrelated to the subject under discussion.}

\item(\textit{Memory and time constraints})
    Human society changes in time. So, first, in studying the regularities of social systems reproducing the initial conditions could be hampered or even impossible. Second, a priory, it is not clear how long the memory of social objects is. In other words, how long time span should separate events in the past from the present instant in order to ignore their effects \emph{within the most detailed description}. For example, stock markets, where human factor definitely matters, exhibit a long-time memory behavior, namely, time correlations in the volatility of returns are characterized by a power decay (see, e.g, \cite{27,28}). Natural systems, by contrast, are characterized by the reproducibility; under the same initial and external conditions either their dynamics or probabilistic characteristics are identical on all the trails. In this meaning, the history of natural systems does not matter.

\item(\textit{Motivation and value factors}) The human behavior is governed by many motives for achieving individual goals as well as obeys the social and cultural norms. There is, typically, a set of possible strategies of behavior among which a decision maker chooses the appropriate one. In doing so he applies to various value factors that reflect his individual preferences and the social and cultural meaning as well (for a review of these aspects see, e.g., \cite{11prev}). Such notions are just inapplicable to natural systems.

\item(\textit{Information deficiency and breakdown of the explicit means-end relationships}) The decision-making environment involves many factors, external and internal ones, that, on one hand, are hidden, i.e. are not recognizable and controllable in principle for decision makers. On the other hand, these factors affect substantially the dynamics of social systems. Therefore, first, decision makers seldom have perfect information about the choice alternatives and their consequences. Second, if a cause and its effect are separated by a significant time interval it could be difficult to recognize and establish their relationship even within a very thorough analysis \cite{11prev}. For the disciplines studying natural objects the existence of the direct means-end relationships is one of their cornerstones.

\item(\textit{Learning, prediction, and social norms}) To choose an appropriate behavior under the information deficiency of social system states human beings draw on either their own experience or the experience of the society. The former one is gained during the learning process based on predicting the results of their actions. The latter one is aggregated in various social norms of human behavior.  In particular, due to effects of the human prediction the dynamics of a social system is affected substantially not only by its history and the current state, but also by its possible future development existing in the human mind (for discussion of these aspects within the so-called intentional rationality see \cite{Beckert}). Such notions are also inapplicable to natural systems.

\end{itemize}

The aforementioned features enable us to claim that the description of statistical social systems requires individual physical notions and mathematical formalism to be developed. Such quantities like forces, free energy, entropy, temperature, etc. could be inapplicable to social systems or their use requires special consideration even at the phenomenological level. In what follows we will discus these points in detail with respect to a certain specific class of such objects, namely, statistical ensembles of elements whose dynamics is governed by motives for their actions, the evaluation of possible behavior strategies and making the appropriate choice with its further correction. We will call such ensembles systems with motivation. Traffic and pedestrian flows, interacting market agents, as well as in some sense bird flocks and fish schools can be regarded as characteristic examples of the systems under consideration.

However before passing directly to the main subject of the paper let us consider from the general point of view the decision-making process. It plays the essential role in a large variety of social systems and its properties should be taken into account in the mathematical description of any statistical social system \cite{6.3,6.4}.

\section{Decision-making process}\label{sec3}

The classical theory of making decisions is based on the notion of the preference relation and the utility function quantifying this relation (see, e.g., \cite{Savage}). The concept of the perfect rationality assumes the human choice or decision to be determined by the most preferable result meeting the maximum of the utility function. The related theory of making decisions under uncertainty also deals with some utility function aggregating in itself the realization of various environmental conditions in a probabilistic way. However, such an approach encounters obstacles caused by the fundamental properties of human beings described above. For example, the possibility of introducing the preference relation with respect to the final goals seems to be doubtful whereas local aims that are similar in value can be indistinguishable for human beings in making decisions.

In order to overcome these obstacles the concepts of bounded rationality \cite{6prev,7prev} and limited cognition \cite{8prev} have been developed (see also \cite{9prev,10prev,11prev}). In particular, it has been proposed \cite{11prev} that the decision-making process (at least in statistical social systems) should be mainly based on selection of possible behavior strategies rather than final goals. Such a strategy is a certain sequence of local actions, i.e. a collection of steps of achieving subsequent intermediate aims. These strategies are formed in the trial-and-error process and evolve during the adaptation of individuals to the decision-making environment under uncertainty of the information about the social system states. Following \cite{11prev} we will call these strategies heuristics.

These heuristics aggregate and accumulate the information about the previous actions, successful and failed ones. That is way the history of a social system impacts on its dynamics. There are at least two distinct ways of the heuristics formation. The first one is the individual learning, i.e. the process of gaining the knowledge about the successful rules of behavior via the personal experience or the experience of individuals directly related to a given one. In particular, the idea that the individual learning plays the leading role in the heuristics formation has been developed in \cite{12prev,13prev} (see also references therein). The second way deals with the cooperative interaction of many individuals forming large units of human society. It is implemented via the formation of the social norms and cultural values aggregating all the fragments of information about the human society for a rather long time interval. The human societies possess own mechanisms governing the social norms and keeping up the social order (see, e.g., \cite{11prev} and references therein). There are at least two types of models for mechanisms via which the social norms and cultural values arise and evolve. One of them is based on emulating the behavior of the most successful persons, i.e. the social \emph{interdependence via significant others} \cite{16prev}. The other type models, \emph{interdependence via reference groups} \cite{17prev}, go beyond the individualistic level of social interdependence. They relate the social and cultural proclivities of human behavior to some large groups or their typical representatives that have high social rank.

\section{Systems with motivation}

In what follows we will confine our consideration to systems with motivation. Keeping in mind traffic flow, ensembles of pedestrians, etc. as specific examples we will develop the main physical notions and concepts of mathematical formalism that allow for the basic features of human behavior discussed above. In this way we will demonstrate, in particular, the feasibility of overcoming the discrepancy between the disciplines studying natural and social systems.

It should be underlined beforehand, that we do not intend to construct a self-consistent and complete theory of systems with motivation; it goes far beyond the scope the present analysis. Our goal is to consider their main features individually and to formulate the conceptual basis for constructing mathematical models for specific phenomena.

\subsection{Extended phase space}\label{EPS}

At the initial step we have to introduce the notion of the ``phase space'', i.e., a collection of variables $\{w\}$ that completely characterize the state of a given social system at the \emph{current} moment of time $t$. Because of the properties listed above the phase space of social systems can differ essentially from that of physical objects.

First, due to the human memory these variables taken at the current moment of time are not necessary to specify the system dynamics. Only having been known at all the previous moments of time, these variables determine the system dynamics completely, may be in a certain probabilistic way. In other words, the rates $\{\delta_t w\}$ of time variations in the phase variables should be certain functionals on themselves rather then some functions,
\begin{equation}\label{2}
\xymatrix{
    \{w[t']\}_{t'<t}\quad \ar@{|->}[r] &\quad \{\delta_t w\}\,.
}
\end{equation}
Here the square brackets at the symbol $w$ stand for the function $w$ of the argument $t'$ rather than its value taken at $t'$ and the symbol $\delta_t w$ denotes the time derivative $\dot{w}$ of the corresponding variable if it is continuous one or, otherwise, specifies step-like jumps between its possible values.

Second, as noted above, the dynamics of a social system is governed by two factors, the direct interaction of its elements, where each of them responds to the behavior of the other elements in some way, and individual motives, wishes, trends, etc, that are not explicitly visible to the others. So the phase space $\{w\}$ of a system with motivation is to comprise variables of two types, objective and subjective ones, $\{w\}=\{q,h\}$, allowing us to call $\{w\}$ the extended phase space. Let us discuss these types of phase variables separately.

\textit{Objective phase space} of an element $\alpha$, by definition, is a collection of variables $\{q\}_\alpha$, discrete or continuous ones, that completely characterize the possible states of the given element $\alpha$ from the standpoint of the other elements. The information about the state of the element $\alpha$ is necessary for them to make the appropriate decisions in governing their own states.  We have used the term ``objective'' in order to underline that the characteristics $\{q\}_\alpha$ of the element $\alpha$ are detectable for the other elements. They are not related to the intentions, plans, wishes of the element $\alpha$ which are hidden for external observers. It should be noted that the variables $\{q\}_\alpha$ are accessible for external observers only in principle. As noted above in a social system getting information about the states of its elements can be hampered. The combination of the objective phase variables of all the elements, $\{q\}$, makes up the objective phase space of the given system. For example, the spatial coordinates of pedestrians, the direction of motion, and may be their velocities form the objective phase space of the pedestrian ensemble, the coordinates and the velocities of vehicles on a highway make up the objective phase space of traffic flow. The set of personal opinions makes the objective phase space of voting process, cultural features with preferences ascribed to every individual can be regarded as the objective phase space of the cultural dynamics. The production and comprehensive matrices characterizing the frequency of using and associating words to the corresponding objects by individuals can be considered in the same way in describing the evolution of languages (see \cite{1} and references therein).

On one hand, the objective phase variables specify the state of a given system from the standpoint of external observer. On the other hand the decision-making, individual motives, wishes, goals, plans etc. rooted in the human mind plays the key role in the dynamics of social systems \cite{6.3,6.4}. So there should be some way for such internal processes to affect the system dynamics. Its implementation for the systems under consideration is related to the notion of \text{controllable} variables forming a special group of the objective phase variables. Elements of social systems try to control their own states via maintaining or changing the objective phase variables or a certain group of them enabling this action directly. The feasibility of this action actually classifies a phase variable as controllable one. The subscript $c$ will be added to the corresponding variables $\{q_c\}$ to underline the given feature. Time variations of the remaining quantities are determined by these controllable variables and, may be, some natural regularities.  For example, in traffic flow a driver can change directly only the velocity of his car. Therefore the velocities of cars are the controllable variables, whereas the coordinates of their position on highways are not so.

Keeping in mind traffic flow and a set of voters as examples of systems with motivation, it is naturally to assume that if the objective phase space contains together with a variable $w$ also its time derivative $\dot{w}$, then the latter is a controllable phase variable whereas the variable $w$ itself is not it. Otherwise, when only the variable $w$ enters the phase space it should be a controllable variable.

\emph{Subjective phase space:} The other part of the phase space, the subjective phase variables, is related to the internal processes in the decision-making. By definition, the subjective variable $\{h\}$ describe time variations in the controllable phase variables, namely, $\{h:=\delta_t q_c\}$. The collection of quantities $\{h\}_\alpha$ ascribed to a given element $\alpha$ will be referred to as the subjective (hidden) phase variables of the element $\alpha$ and the combination of all these quantities will be called the subjective phase space of the given system.

The quantities $\{h\}_\alpha$ characterize active behavior of the element $\alpha$ in governing its state and are related to its motives, wishes, goals, etc. in making decisions. So they are accessible only for the element $\alpha$ and hidden for the others. For every element $\alpha$ its subjective phase variables $\{h\}_\alpha$ are valuable in their own right. This is due to the fact that internal processes accompanying the decision-making themselves take effort in order, for example, to get a decision of changing the current state of the element $\alpha$ \cite{11prev}. In addition, time variations in the quantities $\{q_c\}$ can affect this element in some physical way. For example, the acceleration (or deceleration) of a car can itself affect the stability of car motion and is felt explicitly by the drivers.

\begin{figure}[th]
\centerline{\psfig{file=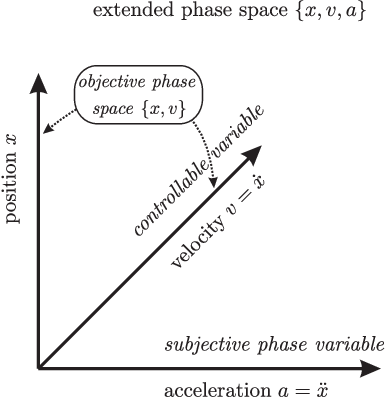,width=5cm}}
\vspace*{8pt}
\caption{Illustration of the structure of the extended phase space for systems with motivation.\label{Fig1}}
\end{figure}

Concluding the present subsection we note once more that the dynamics of systems with motivation is determined directly by the objective and subjective variables simultaneously. That is why the phase space of systems with motivation is made up of both the types of the phase variables, $\{w\} = \{q,h\}$. For example, the accelerations of cars moving on a highway have to be treated as the subjective phase variables of traffic flow and the functions quantifying the quality of individual car motion should contain the car acceleration in the list of their arguments \cite{we1,we2,we3}. Figure~\ref{Fig1} illustrates the structure of the corresponding phase space.

The introduced phase variables form the required language for describing the dynamics of systems with motivation.

\subsection{Decision-making and heuristics choice}

The decision making process governs time variations of the controllable objective variables, which in turn via physical regularities determines the system dynamics as a whole. Symbolically we write this in terms of time increment in the phase variables
\begin{equation}\label{1}
\begin{split}\xymatrix{
    \text{decision-making}\ar@{:>}[d] & \{w\} = \{q,h\}\ar@{.>}[d]
\\
     \{h=\delta_t q_c\}\ar@{:>}[r] &\bigoplus\ar@{.>}[d] & \text{natural regularities}\ar@{:>}[l]
\\
                                   &\{\delta_t w\}
}
\end{split}
\end{equation}
As noted in Sec.~\ref{sec3}, the decision-making is reduced to the choice of local heuristics because of the bounded capacity of human cognition and the variety of factors uncontrollable and hidden for elements of a social system. These heuristics, i.e., the local strategies of the element behavior are sequences of actions focused on achieving local aims. Since in statistical social systems the explicit means-end relationships can be broken the specific actions of elements are evaluated by local motives rather than intentions of getting final goals. The latter goals can only single out some rather general class of the element actions. Moreover the final goals are typically stated in a general form without particular details.

As the systems with motivation are concerned, the heuristics can be regarded as possible trajectories of the forthcoming system dynamics which exist in the human mind and are the unit elements in the decision-making. Therefore at the first step in specifying the heuristics choice we introduce an imaginary phase space $\{\varpi\}_\alpha$ in addition to the real one $\{w\}$ which is ascribed individually to each element $\alpha$. So, in fact, we have introduced the set of spaces existing in the human mind. It is the imaginary phase spaces that enable us to describe a hypothetical dynamics of the system in the near feature expected by its elements based on the available information. Every imaginary phase space
\begin{equation}\label{ips}
   \{\varpi\}_\alpha = \{\boldsymbol{\theta},\eta\}_{\alpha}
\end{equation}
comprises the objective variables $\{\boldsymbol{\theta}\}_\alpha$ of all the elements and the subjective variables $\{\eta\}_\alpha$ of the given element $\alpha$. These quantities specify the hypothetical states of the elements in the ``mind'' of the element $\alpha$. Therefore the symbol $\boldsymbol{\theta}_{\alpha':\alpha}$ is written in bold to underline its dependence on two indices, meaning the phase variables of the element $\alpha'$ in the ``mind'' of the element $\alpha$. Collection~\eqref{ips} does not contain the subjective phase variables of the other elements because they are hidden for the given element $\alpha$.

In these terms a possible strategy of behavior of the element $\alpha$ is represented as a certain time dependence $\{\eta[t'']\}^{t''>t}_\alpha$ of its subjective phase variables in the near future. The hypothetical time dependence $\{\boldsymbol{\theta}[t'']\}^{t''>t}_{\alpha:\alpha}$ of its objective variables is determined by the given strategy of behavior. The hypothetical time dependence $\{\boldsymbol{\theta}[t'']\}^{t''>t}_{\alpha':\alpha}$ of the objective variables ascribed to another element $\alpha'\neq\alpha$ is constructed in the ``mind'' of the element $\alpha$ based on the available information. We also will use the notation $\{\varpi[t'']\}^{t''>t}_\alpha$ to denote this strategy as the hypothetical motion of the system in the space $\{\varpi\}_\alpha$. The symbol $\{\varpi[t'']\}^{t''>t}$ without the element index stands for the heuristics as whole.

\begin{figure}[th]
\centerline{\psfig{file=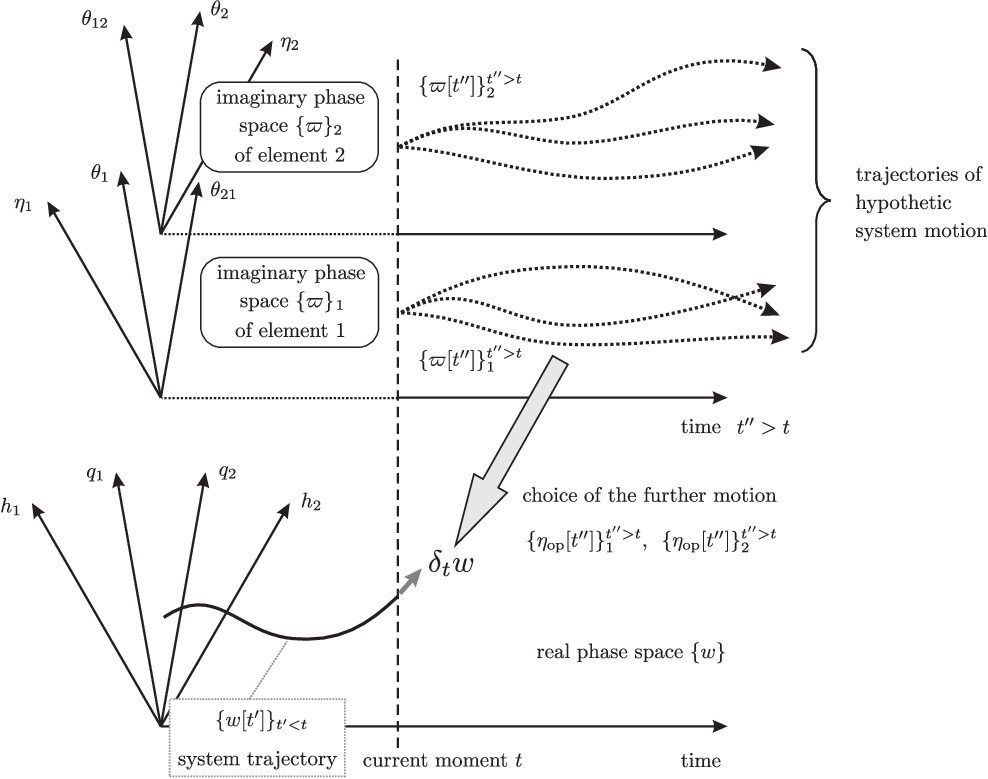,width=\textwidth}}
\vspace*{8pt}
\caption{Illustration of the decision-making process governing the system dynamics.\label{Fig2}}
\end{figure}

The elements are assumed to evaluate and choose the desired strategies of behavior $\{\varpi_\text{op}[t'']\}^{t''>t}$ in some way which determines the system dynamics. In this choice human beings can follow some rational reasons, including social and cultural norms, or even irrational motives, the latter, however, will not be considered in the present paper. It should be noted that in this choice every element $\alpha$ evaluates possible strategies of its own behavior $\{\varpi_\text{op}[t'']\}^{t''>t}_\alpha$ only, the behavior of the other elements is regarded by it as given beforehand or predictable with some probability. These features of the heuristics choice enable us to represent symbolic expression~\eqref{2} as
\begin{equation}\label{3}
\begin{split}
\xymatrix{
   \{w[t'],\varpi[t'']\}^{t''>t}_{t'<t}
    \ar@{=>}[rr] ^-{\text{individual choice}}_-{\text{of system elements}}
    \ar[dr]
    && \ar@{=>}[dl]^{\quad\{h[t'']\}^{t''>t}} \{\varpi_\text{op}[t'']\}^{t''>t}\\
    & \{\delta_t w\}
}
\end{split}
\end{equation}
It should be pointed out that in choosing the heuristics the elements can predict the system dynamics extrapolating the time variations of the phase variables in some simple way, for example, fixing them or supposing the linear time dependence to hold in the near future. Figure~\ref{Fig2} illustrates this.

Concluding the present subsection we state that the laws governing systems with motivation should be based on some variational principles dealing with trajectories $\{\eta[t'']\}^{t''>t}$ in the imaginary subjective spaces of the corresponding elements. As a result the governing equations have to belong to a certain class of temporally boundary value problems because these trajectories join the current state of the system with ``desirable'' ones (cf. \cite{we2}). Some implementation of these variational principles gives us the time variations $\{\delta_t w\}$ of the phase variables at the current time $t$. In this sense the ``imaginary future'' of the system affects its dynamics at present.

\subsection{The approximation of perfect rationality}

The implementation of the variational principles mentioned in the previous subsection requires some measure for quantifying the heuristics with respect to their value at least approximately. This measure can be constructed in a certain limit case called the perfect rationality. It comes into being when, first, analyzed situations are repeated many times, with the environment conditions being the same. Thereby the time restrictions affecting the decision-making process are removed and the complete information about the system becomes accessible. Second, the elements are able to correct their states continuously.

Under such conditions the individual choice of the optimal heuristics $\{\eta_\text{op}[t'']\}^{t''>t}_\alpha$ by a given element $\alpha$ is reduced to finding the maximum of a certain preference functional
\begin{subequations}\label{4all}
\begin{equation}\label{4}
     \mathcal{U}_\alpha := \mathcal{U}_\alpha\Big\{\{\varpi[t'']\}^{t''>t}_\alpha \Big\}
\end{equation}
with respect to its own strategy of behavior $\{\eta[t'']\}^{t''>t}_\alpha$. In the limit of perfect rationality functional~\eqref{4} depends only on the trial trajectory of the system motion in the imaginary space $\{\varpi[t'']\}^{t''>t}$. Besides, all the objective parts of the individual imaginary spaces are identical. The following expression
\begin{equation}\label{4a}
     \mathcal{U}_\alpha := \int\limits_t^{+\infty} dt''e^{-\tfrac{(t''-t)}{T}}u_\alpha(\varpi_\alpha[t''])
\end{equation}
\end{subequations}
is an example of functional~\eqref{4}, where all the moments of time contribute independently of each other with the weight $\exp\{-(t''-t)/T\})$ decreasing exponentially as the time interval $t''-t$ increases, the scale $T$ specifies the temporal horizon of predicting the system dynamics, and the function $u_\alpha(\varpi_\alpha[t''])$ measures the contribution of individual time moments.

Functional~\eqref{4all} quantifies the preferences of the element $\alpha$ in the choice of its own heuristics, provided the behavior of the other elements is known. In other words, within the frameworks of the perfect rationality the optimal strategy of behavior $\{\eta_\text{op}[t'']\}^{t''>t}_\alpha$ is determined by the expression 
\begin{equation}\label{5}
  \{\eta_\text{op}[t'']\}^{t''>t}_{\alpha}
  \quad\Longleftarrow\quad
  \max_{\{\eta[t'']\}^{t''>t}_\alpha}
    \mathcal{U}_\alpha\,.
\end{equation}
Expressions~\eqref{5} specifies the optimal heuristics $\{\eta_\text{op}[t'']\}^{t''>t}$ as certain trajectories. Therefore, in particular, if the elements choose these optimal strategies of behavior at the current moment of time and follow them, then further correction of the system motion will be not necessary. The latter is the essence of the Nash equilibrium.

By way of example, we note that for traffic flow schema~\eqref{3} meeting condition~\eqref{5} in the limit of perfect rationality gives rise to Newtonian type models \cite{we2}. It is the case where the concept of social forces \cite{6} holds. If the driver behavior is not perfect then the description of traffic dynamics goes beyond the notions of Newtonian mechanics \cite{we3}.

The limit of perfect rationality has been included in the present paper to illustrate an way of constructing a measure of the heuristics to be used further. Nevertheless, even in this limit equation~\eqref{5} does not lead directly to the initial value problem, i.e. the class of of mathematical models involving Newtonian mechanics. Indeed, let us assume, for example, that a given system is characterized by continuous objective variables $\{q\}$, all of them being controllable, so $h_\alpha = dq_\alpha/dt$. Then the maximization procedure~\eqref{5} gives us the governing equation in the form
\begin{equation}\label{5a}
   \frac{\partial u_\alpha}{\partial q_\alpha} +\frac1T\frac{\partial u_\alpha}{\partial h_\alpha} -
   \frac{d}{dt''}\frac{\partial u_\alpha}{\partial h_\alpha} = 0\,.
\end{equation}
%
In this case the found governing equation~\eqref{5a} is of the second order with respect to the objective phase variables, whereas the initial conditions specifying the initial system state can contain only the objective variables and, thereby, does not determine uniquely the system dynamics. By this rather simple example we have demonstrated the fact that some terminal conditions should be imposed on the system to determine its dynamics. Therefore we again get the conclusion that the dynamics of the analyzed systems belong to a certain class of temporal boundary value problems. Another fact demonstrated by this example is the influence of the prediction horizon $T$ on the system dynamics. If the parameter $T$ is rather small then the governing equation~\eqref{5a} is practically of the first order with respect to the objective variables $\{q\}$ and in this limit the system dynamics can be considered to be some initial value problem. In the opposite limit, i.e. when the parameter $T$ is large the effect of the terminal conditions is crucial.

\emph{Rational dynamics attractor:} In order to discuss further a new type of nonequilibrium phase transitions caused by the bounded capacity of the human cognition let us single out a special subclass of systems with motivation. For these systems the governing equation~\eqref{5} admits the steady state solution corresponding to the union of the origin $\{\eta = 0\}$ of the subjective phase space and a certain point $\mathfrak{Q}$ (or a set of points with equal values of the controllable variables, $\{q_c = \text{const}\}$) of the objective phase space. If the system is initially located at this point and its elements do nothing  with respect to controlling its dynamics, i.e., keep up the current values of the controllable variables, then it will not leave this point further. For example, traffic flow where all the cars move with the same speed and at some optimal headway distance matches this situation. If the system during its motion governed by expression~\eqref{5} tends to the point $\mathfrak{Q}$ or the corresponding set of points, it will be referred to as an attractor of rational dynamics.

\subsection{Bounded rationality and action points}

In this section we returns to the main subject and discuss the basic notions of the elementary dynamics that a necessary for describing systems with motivation. As noted previously, the time constraints together with the bounded capacity of human cognition endow the choice of heuristics and, thereby, the system dynamics with random properties. If two strategies of behavior are rather close to each other in value then it can be tough to order them by preference and to choose one in a rational way. To tackle this problem we appeal to the notion of perception threshold $\Theta$. The perception threshold as well as the preference functional depends generally on the type of elements, which here is not labeled directly to simplify the notations.

\emph{Bounded rationality:} Let us make use of the preference functional~\eqref{4}. Two strategies of behavior $\{\eta_1[t'']\}^{t''>t}_\alpha$ and $\{\eta_2[t'']\}^{t''>t}_\alpha$ are considered to be equivalent, if the corresponding magnitudes of the preference functional meet the inequality $\left|\mathcal{U}_{1,\alpha}-\mathcal{U}_{2,\alpha}\right| \lesssim \Theta$, provided the other environment conditions are the same. It is the point where the form of the preference functional~\eqref{4} becomes determined. The matter is that the action of any increasing function on the preference functional gives rise to a new preference functional describing the same set of the optimal heuristics. Introducing the perception threshold we actually fix its form.

When a currently chosen strategy of behavior $\{\eta[t'']\}^{t''>t}_\alpha$ is close to the optimal one in the given sense, the element $\alpha$ has no motives to change it. Roughly speaking, if it is not clear what to do, to change nothing is quite adequate. If the difference in the magnitudes of the preference functional~\eqref{4} for the two strategies becomes remarkable in comparison with the perception threshold, the element $\alpha$ recognizes the necessity of correcting its current state. Exactly this choice of new more proper heuristics is the point where the random factors enter the system directly. Indeed, since all the strategies of behavior that are close to the optimal heuristics in terms of the perception threshold are regarded as equivalent then the choice of some of them is a random event. The time moment when this choice arises is also a random quantity.

It should be pointed out that the perception threshold $\Theta$ characterizes probabilistic properties of the element behavior rather than the step-like dynamics. Namely, let us consider two heuristics, the strategy of behavior $\{\eta_c[t'']\}^{t''>t}_\alpha$ that is taken by the element $\alpha$ at the current moment of time $t$ and the optimal one $\{\eta_\text{op}[t'']\}^{t''>t}_\alpha$ which is actually hidden for it. When the difference in the corresponding magnitudes of the preference functional~\eqref{4} becomes equal to the threshold, $|\mathcal{U}_{c,\alpha} -\mathcal{U}_{\text{op},\alpha}| = \Theta$, the events of correcting the current state by the element $\alpha$ just arise most offen rather than exhibits a stepwise behavior. In the cases $|\mathcal{U}_{c,\alpha} -\mathcal{U}_{\text{op},\alpha}| \ll \Theta$ the element cannot recognize the fact of the system deviating from the optimal dynamics. States matching the opposite inequality $|\mathcal{U}_{c,\alpha} -\mathcal{U}_{\text{op},\alpha}| \gg \Theta$ cannot be reached because the element would respond earlier. In particular, according to empirical data for traffic flow such events of correcting the car motion are distributed rather widely near the corresponding threshold \cite{24prev}.

\textit{Action points:}
Let us introduce the notion of action points in order to describe the dynamics of systems with motivation. An action point is an event of changing the current strategy of behavior by some element in correcting its state. Every action point is associated with this strategy and the time moment of changing it. When the dynamics of a given system is optimal within the human perception characterized by the threshold $\Theta$ its elements do not correct their heuristics. At these moments of time the system motion is not controlled by the elements and proceeds according to natural regularities affecting the system and the strategy chosen previously. When the system motion deviates from the optimal one substantially the elements recognize this fact and correct their individual strategies of behavior. In doing so an element selects some new strategy of behavior in a neighborhood of the optimal heuristics whose thickness quantified by the preference functional~\eqref{4} is less than or of the order of the perception threshold. Then the given element follows the selected heuristics until it recognize the necessity of correcting its state again. We note that the notion of action points for the car-following process was introduced for the first time in \cite{23} to denote the moments of time when drivers correct the motion of their vehicles by pressing the gas or braking pedals.

\begin{figure}[th]
\centerline{\psfig{file=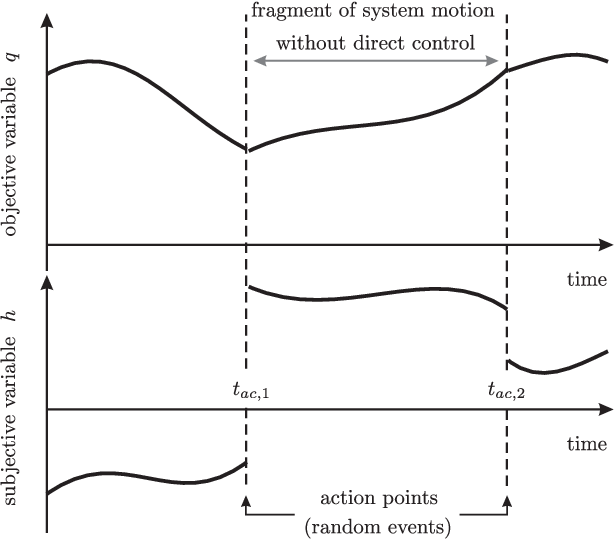,width=0.7\textwidth}}
\vspace*{8pt}
\caption{Illustration of the system dynamics governed by elements with bounded rationality.\label{Fig3}}
\end{figure}

In these terms the dynamics of a system with motivation can be represented as a sequence of action points, i.e. jumps between various strategies of the element behavior. The particular strategies of behavior joined by these jump-like transitions and their time moments are random quantities. These random transitions are the cause of the stochasticity in the dynamics of systems with motivation. Between the action points the system dynamics is not controlled by its elements at all and is regular or affected by random factors of natural origin. Symbolically this feature is represented by the following diagram generalizing the previous one~\eqref{3}

\begin{equation}\label{6}
\begin{split}\xymatrix{
   \{w[t'],\varpi[t'']\}^{t''>t}_{t'<t}
    \ar@{=>}[rr] ^-{\text{individual choice}}_-{\text{of system elements}}
    \ar@{->}[d]
    && \ar@{=>}[dl]^{\text{\quad action points}} \{\varpi_\text{op}[t'']\}^{t''>t}\\
    \{\delta_t w\} & \{\delta_t h\}\ar@{=>}[l] &
}
\end{split}
\end{equation}
The concept of action points is illustrated in Fig.~\ref{Fig3}.

\textit{Dynamical traps:}
If a given system possesses an attractor of rational dynamics it can exhibit a new type of cooperative phenomena. Indeed, by definition,  the point $\mathfrak{Q}$ in the objective phase space matches the steady-state dynamics in the limit of perfect rationality. Then the elements with bounded rationality will regard the system motion in the vicinity of this attractor also optimal. To discuss the given feature in more detail let us introduce the notion of dynamical traps.

\begin{figure}[th]
\centerline{\psfig{file=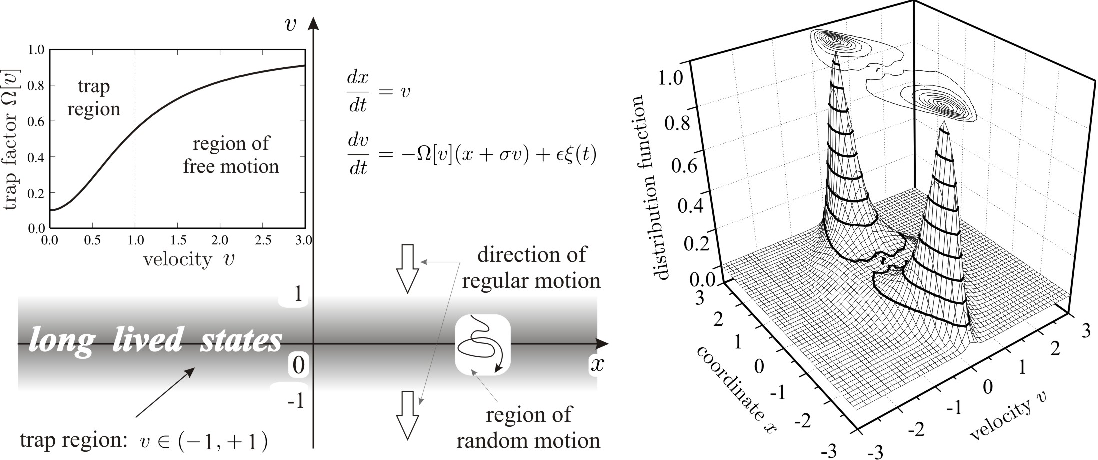,width=\textwidth}}
\vspace*{8pt}
\caption{Illustration of nonequilibrium phase transitions induced by dynamical traps. The figure presents one-dimensional oscillator with dynamical traps affected by white noise $\epsilon\xi(t)$ of a small amplitude $\epsilon\ll1$ and energy dissipation characterized by the friction coefficient $\sigma$. The dynamical trap region is a certain neighborhood of the $x$-axis. The effect of dynamical traps is described by the factor $\Omega[v]$ depending on the velocity $v$; inside the traps region $\Omega[v]\ll1$, outside it $\Omega[v]\approx1$. Therefore outside the region of dynamical traps the system motion is rather regular, whereas inside it the oscillator dynamics is stagnated and affected mainly by weak noise only. As demonstrated numerically \cite{DT2}, the system undergoes phase transition; the form of the distribution function in the space $\{x,v\}$ is converted from unimodal to bimodal one as the intensity of dynamical traps grows. The resulting bimodal distribution is shown in the right fragment of this figure.\label{Fig4}}
\end{figure}

Using the preference functional~\eqref{4} we construct a certain neighborhood of the set $\mathfrak{Q}\bigotimes\{h=0\}$ in the space of heuristics $\{\varpi[t'']\}^{t''>t}$ of thickness $\Theta$ and then project it onto the objective phase space $\{w\}$. In this way we obtain a certain neighborhood $\mathfrak{D_Q}$ of the set $\mathfrak{Q}$ called the dynamical trap region. When the system with bounded rationality enters this region its elements consider the system dynamics optimal and the correction of their state unnecessary. Since the system dynamics in the region $\mathfrak{D_Q}$ is really close to the optimal one, a time span between two action points could rather prolonged in comparison with that of the system dynamics far from $\mathfrak{D_Q}$. In other words such fragments of system motion inside the dynamical trap region can be regarded as long-lived states \cite{DT1,DT2}. Their origin is due to the stagnation of the element active behavior during a relatively long time. Therefore dynamical traps are able to induce nonequilibrium phase transitions of a new type that should be widely met in social systems \cite{DT1,DT2,DT3} rather than in natural ones. We note that the dynamical traps for Hamiltonian systems was introduced in \cite{Zas1,Zas2} (see also a review \cite{ZasRev}) and for systems with nonlinear oscillations it was done in \cite{Gaf}. A simple example of nonequilibrium phase transitions induced by dynamical traps is illustrated in Fig.~\ref{Fig4}.

\subsection{Memory effects: individual learning and formation of social and cultural norms}

In the previous subsections we have considered the description elements of the decision-making related to predicting the forthcoming events. Here the effects of the system history is under consideration.

\emph{Knowledge accumulation:}
Because of the bounded capacity of human cognition, gaining the knowledge about the proper strategies of behavior is crucial. As noted in Sec.~\ref{sec3} there are two channels of accumulating and aggregating such information. One is the individual learning of the elements based on own experience or local interaction with the other elements. In some sense it is a typical mechanism of cooperative phenomena widely met in natural systems and caused by local or quasi-local interaction of their particles. It seems to be possible to describe the individual learning process using the introduced phase space and perception thresholds. No additional variables are necessary to do this. Indeed, let us ascribe an individual perception threshold $\Theta_\alpha$ to every element $\alpha$. Then the individual learning is represented as the evolution of the perception thresholds $\{\Theta_\alpha\}$ caused by some interaction of the elements. Symbolically it takes the form
\begin{equation}\label{77}
    \{w[t']\}_{t'<t} \Longrightarrow \{\Theta_\alpha(t)\}\,.
\end{equation}
In these terms the individual learning is reduced to the time decrease of the perception thresholds $\{\Theta_\alpha\}$ due to the accumulation and aggregation of information about the system properties.

The second channel is related to a unique collective interaction of all the elements in a social system in addition to their individual interrelations of various types. It arises via the formation of social and cultural norms of behavior. These norms affect directly the heuristics and their preference, and involve all the members of a social system or their large groups independently of their relationships and distance in space and time. The social and cultural norms aggregate the information about the properties and features of a social system during a long time interval and make up the basis for finding \emph{general} rules of successful strategies of behavior. So in order to describe the effect of the social and cultural norms on the system dynamics some additional variables, the space of cultural features $\{\chi\}$, should be introduced. We presume that the cultural features cannot be ascribed to individual persons in any way, they have their own carriers, e.g., books, newspapers, magazines, movies, broadcasts, and other types of mass media. Symbolically the formation of social and cultural norms can be written as
\begin{equation}\label{7}
    \{w[t']\}_{t'<t} \Longrightarrow \{\chi(t)\}\,,
\end{equation}
where, as it is rather natural to assume, the father a \emph{given} event in the past, the weaker its influence on the present. In order to include the effect of these norms on the social system dynamics we generalize diagram~\eqref{6} as follows
\begin{equation}\label{8}
\begin{split}\xymatrix{
   \{w[t']\}_{t'<t} \ar@{=>}[r] \ar@{~>}[rd]_{\text{individual learning\qquad}}
      & \text{norm space }\{\chi(t)\} \ar@{~>}[d]
\\
   \{w[t'],\varpi[t'']\}^{t''>t}_{t'<t}
    \ar@{=>}[rr] ^-{\text{individual choice}}_-{\text{of system elements}}
    \ar@{->}[d]
    &{\protect\phantom{\Big(sssssss\Big)}}& \ar@{=>}[dl]^{\text{\quad action points}} \{\varpi_\text{op}[t'']\}^{t''>t}
\\
    \{\delta_t w\} & \{\delta_t h\}\ar@{=>}[l] &
}
\end{split}
\end{equation}
which is the final diagram presenting the essence of the mathematical components and notions describing the systems with motivation.

\emph{Problem of initial conditions:}
As noted in Sec.~\ref{sec:pecul} the memory effects and time constraints pose a question as to whether the notion of initial conditions are applicable at all to the statistical social systems. In the remaining part of the paper we will discuss this problem considering as a particular example the car following dynamics.

Let us confine ourselves to describing the dynamics of a car following a lead car moving, e.g., at a fixed speed $V$ in the spirit of the social force model. The social force model \cite{6} relates the acceleration $a$ of the following car to its current velocity $v$ and the distance between the two cars (headway distance) $h$,
\begin{equation}\label{sfm1}
   a = \mathcal{F}(h,v)\,.
\end{equation}
Hear $\mathcal{F}(h,v)$ is a certain function which takes into account two stimuli in car driving, the necessity of maintaining safe headway and zero value relative velocity. It should be underlined that model~\eqref{sfm1} deals with the acceleration $a$ and velocity $v$ of the following car as well as the headway distance $h$ taken at the current moment of time $t$. However, because of the bounded capacity of human cognition drivers cannot recognize the necessity to correct the car motion immediately. The information about the state of motion should be aggregated and accumulated during some time for a driver to make the proper conclusion about correcting its motion. Keeping in mind expressions similar to \eqref{77} and \eqref{7} let us represent the relationship between the acceleration $a$ and the social force $\mathcal{F}(h,v)$ in the following functional form
\begin{equation}\label{sfm20}
   a(t) = \int\limits_{-\infty}^t dt'K(t,t')\,\mathcal{F}\left(h[t'],v[t']\right)
\end{equation}
with a kernel $K(t,t')$ decreasing as the analyzed point in the past, $t'$, goes away from the current moment of time $t$.

Within a simple model of the driver memory characterized by the time scale $T$ the kernel of the integral operator~\eqref{sfm20} is approximated by the exponential function $K(t,t') = 1/T\cdot \exp[-(t-t')/T]$. This enables us to rewrite \eqref{sfm20} as
\begin{equation}\label{sfm200a}
   a(t) = \int\limits_{-\infty}^t \frac{dt'}{T}\, e^{-\tfrac{(t-t')}{T}}\,\mathcal{F}\left(h[t'],v[t']\right)\,.
\end{equation}
or using the properties of the exponential function to represent \eqref{sfm200a} in the form
\begin{equation}\label{sfm200b}
   a(t) = \int\limits_{-t_0}^t \frac{dt'}{T}\, e^{-\tfrac{(t-t')}{T}}\,\mathcal{F}\left(h[t'],v[t']\right) + e^{-\tfrac{(t-t_0)}{T}}a_0\,,
\end{equation}
%
where
\begin{equation}\label{sfm200c}
   a_0 = \int\limits_{-\infty}^{t_0} \frac{dt'}{T}\, e^{-\tfrac{(t_0-t')}{T}}\,\mathcal{F}\left(h[t'],v[t']\right)
\end{equation}
is the car acceleration taken at a certain time moment $t_0<t$ chosen arbitrarily. The integral equation~\eqref{sfm200b} is equivalent to the following differential equation
\begin{equation}\label{sfm201}
   \frac{da}{dt} = \frac1T\mathcal{F}\left(h,v\right) - \frac{a}{T}\,.
\end{equation}
subjected to the initial condition specified by expression~\eqref{sfm200c}
\begin{equation}\label{sfm201in}
   a|_{t=t_0} =  a_0\,.
\end{equation}
Therefore for the exponential kernel of the integral Volterra equation~\eqref{sfm20} the description of memory effects in the car following is reduced to a certain differential equation determined in the expanded phase space containing the car acceleration as a phase variable. This differential equation can be subjected to the corresponding initial condition at an arbitrary moment of time.

For the general form of the kernel $K(t,t')$ the conversion of the integral operator~\eqref{sfm20} into a differential equation subjected to some initial conditions becomes impossible. In this case the notion of initial conditions is inapplicable. In systems where human memory and perception play are crucial, it is the case, e.g., for traffic flow, the kernel $K(t,t')$ seems to be a power function rather than the exponential one. In fact, general speculations about the human memory and the event perception prompt us to make use of a scale-free-memory model to describe the effects of the system history. To justify our point of view, let us consider two events characterizing the system state in a similar manner, which enables us to compare them with each other in assessing the current situation. If one of the two events happened one day before the current date whereas the other happened one week ago, then we will treat them as substantially different in time with respect to their contribution to our perception of the present situation. By contrast, if the first event occurred one month and one day ago and the second event occurred one month and one week ago we will draw no real distinction between each other by the time of their occurrence in evaluating their significance. In other words, if the time lag between the two events is comparable with the time scale separating them from the present moment then their impacts will be regarded to be different in magnitude with respect to the time of occurrence. On the contrary, if their time lag is much less then the passed time these events can be considered to be simultaneous in evaluating their impacts. Exactly such a behavior is common to power dependencies $K(t-t')\propto (t-t')^{-(1-\gamma)}$ with an the
exponent $0<\gamma <1$. The inequality $\gamma> 0$ has to hold because, otherwise, functional~\eqref{sfm20} would be reduced to a local relationship without memory effects.

This idea is partly justified by the observed long-time memory effects in the scale-free foraging by primates \cite{memory1,memory1a,memory3}  or insects \cite{memory4,memory5}, and the conclusion about the explicit relationship between scale-free foraging and the memory properties \cite{memory6}. The human memory retrieval is also characterized by a scale-free pattern \cite{memory7}. In addition, as noted in Introduction, stock markets, where human factor definitely matters, exhibit a long-time memory behavior of the type under consideration, namely, time correlations in the volatility of returns are characterized by a power decay (see, e.g, \cite{27,28}).

Nevertheless for such systems there is a certain approximation wherein the notion of initial conditions can be introduced \cite{Kanemoto}. It is related to the assumption that within a sufficiently long time interval of duration $T$ in evaluating the action preference the system elements remember the time moments of events when they happened and their contribution to the perception of the current situation is weighted by the the power type kernel $K(t - t')$. On temporal scales larger than $T$ the elements do not rank the events according to the time of their occurrence, they just fix these events in the memory. In addition it is assumed that there is a certain moment of time $t_0$ at which the system dynamics was initiated, for example, when the given group of cars entered the highway under consideration. All the previous events happened before the system initiation are just aggregated without ranking them according to the time of occurrence.  In this case the integral Volterra equation~\eqref{sfm20} is approximated as follows \cite{Kanemoto}
\begin{equation}\label{mb:9final}
     a(x,t) = \int\limits_{t_0}^t dt'\,
     \frac{E_{\gamma,\gamma}\left[-\left(\tfrac{t-t'}{T}\right)^\gamma \right]}{\tau^{\gamma}(t-t')^{1-\gamma}}\, \mathcal{F}\left(h[t'],v[t']\right)
     + E_\gamma\left[-\left(\frac{t-t_0}{T}\right)^{\gamma}\right] \,a_0\,.
\end{equation}
where $\tau$ is a certain microscopic time scale, $a_0$ is the initial car acceleration chosen by the driver at the initial time $t_0$, and  $E_{\gamma}(\ldots)$, $E_{\gamma,\gamma}(\ldots)$ are the Mittag-Leffler functions \cite{Kilbas}. Using the formalism of fractional calculus this integral equation is be reduced to the following fractional differential equation \cite{Kilbas}
\begin{equation}\label{mb:11}
  {}^C\!\widehat{D}_{t_0}^\gamma a(t) = \frac{1}{\tau^\gamma} \mathcal{F}\left(h[t],v[t]\right) - \frac1{T^{\gamma}}\,a(t)\,,
\end{equation}
where the left-hand side is the Caputo fractional derivative of order $\gamma$ defined by the expression
\begin{equation}\label{mb11RL}
   {}^C\!\widehat{D}_{t_0}^\gamma a(t):=
   \frac1{\Gamma(1-\gamma)}\int\limits_{t_0}^t \frac{dt'}{(t-t')^{\gamma}} \frac{da(t')}{dt'} 
\end{equation}
and $\Gamma(\ldots)$ is the gamma function. Equation~\eqref{mb:11} should be subjected to the initial condition~\eqref{sfm201in}. 

In other words, for systems with motivation it could be possible to introduce, at least, approximately the initial conditions at the moment of the system initiation only. In addition, the fractional calculus is likely to be an appropriate formalism for describing effects of the human memory.

\section{Conclusion}

The paper has considered systems with motivation as a typical example of statistical social systems. First, the elements of such a system are characterized by motivated behavior and the decision-making process governs the system dynamics. Second, all the elements can be divided into large groups by their properties and similarity. Therefore the local self-averaging holds in these system, enabling us to introduce the notion of the characteristic elements. Their regular properties describe the common features of the elements, whereas the random ones take into account the individuality of the elements as well as unpredictable factors of their behavior.

The purpose of the paper is to pose the question what physical notions and mathematical formalism should be used or developed to describe the statistical social systems, or more strictly, the systems with motivation. In order to demonstrate that the given question is not trivial up to know we have listed the basic properties of social systems distinguishing them from physical ones or, more generally, natural systems. Not intending to create a self-consistent and complete theory of such systems we have analyzed individually their characteristic features caused by the basic properties of human beings and human societies. In particular, we have demonstrated that modeling the description of these systems requires a extended phase space comprising objective and subjective phase variables, the decision-making process deals with the trajectories of the forthcoming system dynamics as the basic units of description, the action points and dynamical traps are basic notions in describing dynamics of elements with bounded rationality, and the system history can play an essential role.
 
As the final conclusion we state that practically all the basic elements necessary for describing statistical social systems or, more rigorously, systems with motivation are available or partly developed in modern physics and applied mathematics.

The work was partially supported by RFBR Grants 06-01-04005 and 09-01-00736.

\end{document}